\documentstyle[11pt,aaspp4]{article}
\begin{document}
\title{A Definitive Optical Detection of a Supercluster at $z \approx 0.91$}

\author{Lori M. Lubin\altaffilmark{1}, Robert Brunner, and Mark R. Metzger}
\affil{California Institute of Technology, Mail Stop 105-24, Pasadena, CA 91125}

\author{Marc Postman}
\affil{Space Telescope Science Institute\altaffilmark{2}, 3700 San Martin Drive, Baltimore, MD 21218}

\author{J. B. Oke}
\affil{California Institute of Technology, Mail Stop 105-24, Pasadena,
CA 91125 and}
\affil{Dominion Astrophysical Observatory, 5071 W. Saanich Road, Victoria,
BC V8X 4M6}

\vskip 1 cm
\centerline{Accepted for publication in the {\it Astrophysical Journal Letters}}

\altaffiltext{1}{Hubble Fellow}

\altaffiltext{2}{Space Telescope Science Institute is operated by the
Association of Universities for Research in Astronomy, Inc.,
under contract to the National Aeronautics and Space Administration.}

\vfill
\eject

\begin{abstract}

We present the results from a multi-band optical imaging program which
has definitively confirmed the existence of a supercluster at $z
\approx 0.91$.  Two massive clusters of galaxies, CL1604+4304 at $z =
0.897$ and CL1604+4321 at $z = 0.924$, were originally observed in the
high-redshift cluster survey of Oke, Postman \& Lubin (1998).  They
are separated by $4300~{\rm km~s^{-1}}$ in radial velocity and 17
arcminutes on the plane of the sky.  Their physical and redshift
proximity suggested a promising supercluster candidate.  Deep $BRi$
imaging of the region between the two clusters indicates a large
population of red galaxies.  This population forms a tight, red
sequence in the color--magnitude diagram at $(R-i) \approx 1.4$. The
characteristic color is identical to that of the
spectroscopically-confirmed early-type galaxies in the two member
clusters.  The red galaxies are spread throughout the $5~h^{-1}~{\rm
Mpc}$ region between CL1604+4304 and CL1604+4321.  Their spatial
distribution delineates the entire large scale structure with high
concentrations at the cluster centers.  In addition, we detect a
significant overdensity of red galaxies directly between CL1604+4304
and CL1604+4321 which is the signature of a third, rich cluster
associated with this system. The strong sequence of red galaxies and
their spatial distribution clearly indicate that we have discovered a
supercluster at $z \approx 0.91$.

\end{abstract}


\section{Introduction}

Superclusters comprise the largest known systems of galaxies,
containing 2 -- 5 massive clusters and extending over 10 -- 20 Mpc
(e.g., Bahcall \& Soneira 1984; Postman, Geller \& Huchra 1988;
Quintana et al.\ 1995; Small et al.\ 1998).  Since the dynamical
timescales of superclusters are comparable to the Hubble time, large
scale structures observed today are cosmic fossils of conditions that
existed in the early universe.  As a result, studies of these systems
can be used to measure the cosmological density parameter $\Omega_o$,
to constrain the large-scale variation of the mass-to-light ratio, and
to test theories of the formation and evolution of galaxies and
clusters (e.g., Hoffman et al.\ 1982; Shaya 1984; Peebles 1986; Cen
1994).  Large scale structures have been studied at low redshift via
the local, Shapley, and Corona Borealis superclusters (e.g., Davis et
al.\ 1980; Postman et al.\ 1988; Quintana et al.\ 1995; Small et al.\
1998).  At higher redshifts of $z \approx 0.4$, weak lensing studies
of MS0302+16 (Kaiser et al.\ 2000) provide a direct measure of the
projected mass distribution on 10 Mpc scales. These studies indicate
that supercluster masses are $M \sim 10^{16} - 10^{17}~h^{-1}~{\rm
M_{\odot}}$, their mass-to-light ratios are $M/L_{B} \sim 200 -
600~h~M_{\odot}/L_{\odot}$, and the density parameter measured on
supercluster scales is $\Omega_{o} \sim 0.1 - 0.4$.  This work can be
extended to higher redshift, but until recently no such systems were
known.

\begin{figure}
\plotone{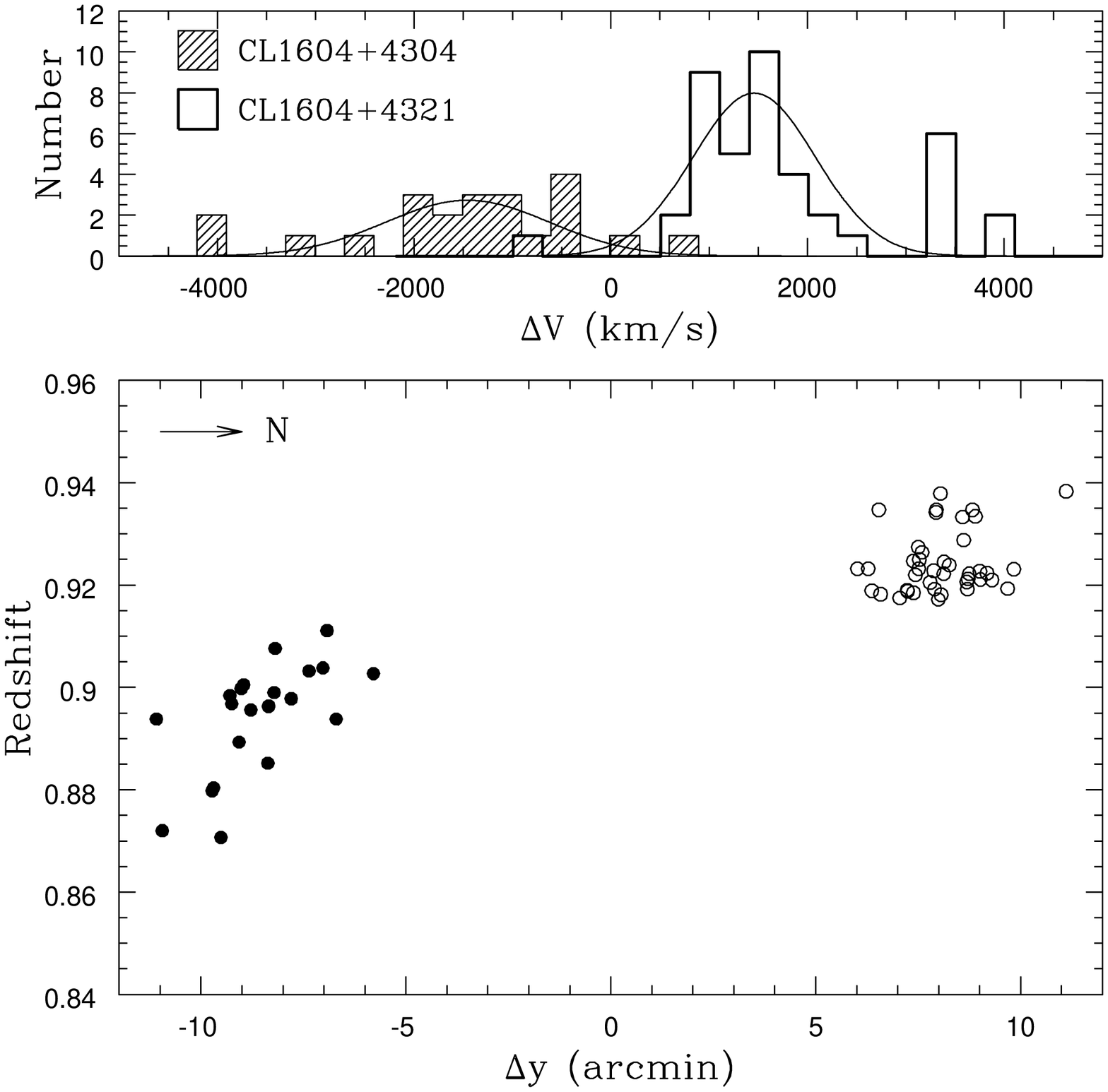}
\caption{{\it Upper Panel} -- Histogram of the
relativistically-corrected velocity offsets for the 63 confirmed
cluster members of CL1604+4304 and CL1604+4321. The offsets are
calculated relative to the mean velocity of the supercluster. The
solid lines indicate the best-fit Gaussian distribution for each
cluster (Postman, Lubin \& Oke 1998, 2000).  {\it Lower Panel} -- The
north--south sky positions versus redshift of the confirmed cluster
members in CL1604+4304 (filled circles) and CL1604+4321 (open
circles).  Though the southern cluster CL1604+4304 has a considerably
larger scatter in velocity, there is a clear trend in which the
redshift on the north side of CL1604+4304 approaches the redshift of
the northern cluster CL1604+4321.}
\label{sc}
\end{figure}

Such a high-redshift system has been discovered in a study of nine
candidate clusters at $z \ga 0.6$ by Oke, Postman \& Lubin (1998).
Two clusters, CL1604+4304 at $z = 0.897$ and CL1604+4321 at $z =
0.924$, were observed as part of this survey.  They are separated by
$4300~{\rm km~s^{-1}}$ in radial velocity and by 17 arcminutes on the
sky. This implies a projected separation of only $5~h^{-1}~{\rm Mpc}$.
We have already analyzed the spectra of a nearly complete sample of
galaxies with $R \le 23.5$ in a $2.2^{'} \times 7.2^{'}$ field
centered on each cluster. The top panel in Figure~\ref{sc} shows the
combined velocity histogram of the 63 confirmed members in the two
clusters (21 in CL1604+4304 and 42 in CL1604+4321). CL1604+4304 and
CL1604+4321 have velocity dispersions of $1226^{+245}_{-154}$ and
$935^{+126}_{-91}~{\rm km~s^{-1}}$ and masses of $3.1$ and $1.6 \times
10^{15}~h^{-1}~{\rm M_{\odot}}$, respectively (Postman, Lubin \& Oke
1998, 2000).  All of the observational data suggest that CL1604+4304
and CL1604+4321 are typical of Abell richness class 1 to 3 clusters
(Lubin et al.\ 1998, 2000).

More interestingly, these two clusters may comprise an even larger
system of galaxies. The lower panel of Figure~\ref{sc} shows the
north-south position of the confirmed cluster members versus redshift.
There is a clear trend in which the redshift on the north side of
CL1604+4304 approaches the redshift of CL1604+4321. The apparent
alignment in redshift space and the physical proximity of the clusters
indicate that this may be a high-redshift supercluster.  The estimated
mass of this structure is $\ga 5 \times 10^{15}~{\rm M_{\odot}}$, and
the spatial overdensity is $\sim 40$.  These numbers imply that the
system is bound and has likely reached turnaround for reasonable
cosmologies (Small et al.\ 1998).

In this Letter, we provide new evidence from multi-band optical
imaging which strongly favors the supercluster hypothesis.  Unless
otherwise noted, we use $H_{0} = 100~h~{\rm km~s^{-1}~Mpc^{-1}}$ and
$q_{0} = 0.1$.

\section{The Observations}

All of the optical imaging was completed with the Carnegie
Observatories Spectroscopic Multislit and Imaging Camera (COSMIC;
Kells et al.\ 1998) at the 200-in Hale telescope at Palomar
Observatories.  We have used the instrument in direct imaging mode
which provides a pixel scale of 0\farcs{285} per pixel and a
field-of-view of $9.7^{'} \times 9.7^{'}$.  Two individual pointings
were made in order to cover the region between CL1604+4304 and
CL1604+4321. Each pointing covered a portion of one cluster.  The
overlap between the pointings was $35^{''}$.  The photometric survey
was conducted in three broadband filters $B$, $R$, and Gunn $i$.  The
total integration times on each pointing were 2 hours in $B$ and 1
hour in $R$ and $i$.  The data were calibrated to the standard
Cousins-Bessell-Landolt system through exposures of Landolt
standard-star fields (Landolt 1992). Variations about the nightly
photometric transformations are 0.03 mag or less.

Source detection and photometry were performed using SExtractor
version 2.1.0 (Bertin \& Arnouts 1996). SExtractor was chosen for its
ability to detect objects in one image and analyze the corresponding
pixels in a separate image. When applied uniformly to multi-band data,
this technique generates a matched aperture dataset. Our detection
image was constructed from the $BRi$ images using a $\chi^2$ process
(Szalay, Connolly \& Szokoly 1998). Briefly, this process involves
convolving each input image with a Gaussian kernel matched to the
seeing. The convolved images were squared and normalized so that the
background had zero mean and unit variance. The three processed images
(corresponding to the original $BRi$ images) were coadded, forming the
$\chi^2$ detection image. A histogram of the pixel distribution in the
$\chi^2$ image was created and compared to a $\chi^2$ function with
three degrees of freedom (which corresponds to the sky pixel
distribution). The difference between the actual pixel distribution
and the $\chi^2$ function provides an optimal estimate for the actual
object pixel distribution. The Bayesian detection threshold was set
equivalent to the intersection of the ``sky'' and ``object''
distributions (i.e.\ where the object pixel flux becomes dominant). To
convert this empirical threshold for use with SExtractor, we scale the
threshold (which is a flux per pixel value) into a surface brightness
threshold (which is in magnitudes per square arcsecond) by defining a
detection zeropoint using the desired detection threshold and the
pixel scale.  Approximately 4800 galaxies were detected in the
combined fields.

For the color analysis, we use the total magnitudes as calculated by
SExtractor.  These magnitudes are variable-diameter aperture
magnitudes measured in an elliptical aperture of major axis radius $2
\times r_k$, where $r_k$ is the Kron radius (see Bertin \& Arnouts
1996).  Because we have used a matched aperture analysis, the total
magnitudes in the three bands are measured within the same physical
radius for a given galaxy.  The limiting magnitudes of our survey are
$B = 25.8$, $R = 24.6$, and $i = 23.5$ for a $5 \sigma$ detection.

\section{The Results}
\subsection{The Galaxy Colors}

\begin{figure}
\plotone{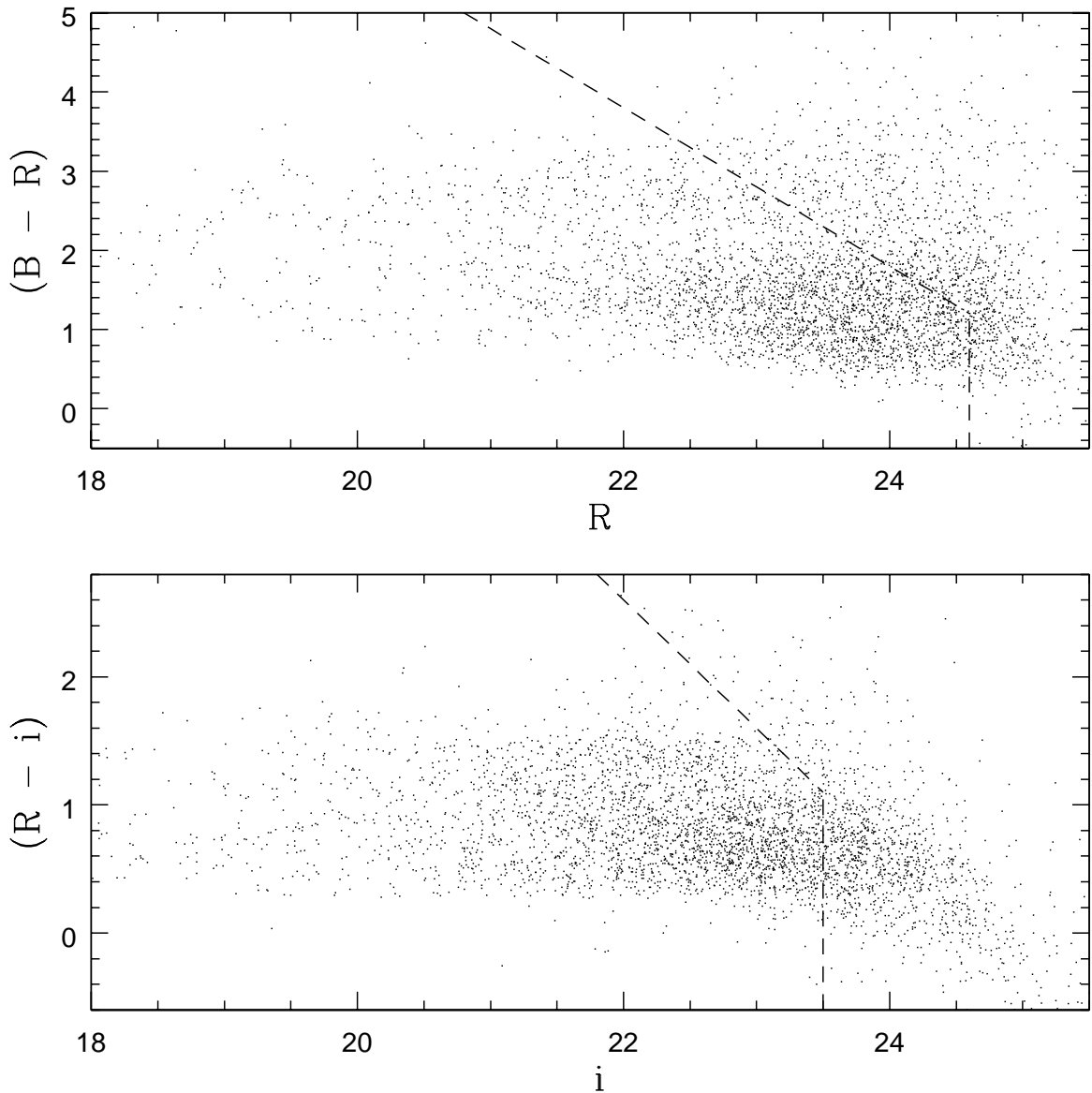}
\caption{The resulting $(B-R)$ and $(R-i)$ color--magnitude (CM)
diagrams from the Palomar imaging of the supercluster region. The
small dots represent all galaxies which have been detected in the
images. The dashed lines indicate the magnitude limits of the
survey. The CM diagrams show a well-defined sequence of red galaxies,
most notably in the $(R-i)$ CM diagram.  In the $(B-R)$ CM diagram,
the larger color scatter in the red sequence is a result of the fact
that many of these galaxies are at or beyond the magnitude limits of
this survey. The presence of this red sequence over the entire
supercluster region strongly supports the existence of a large scale
structure.}
\label{cm} 
\end{figure}

\begin{figure}
\plotone{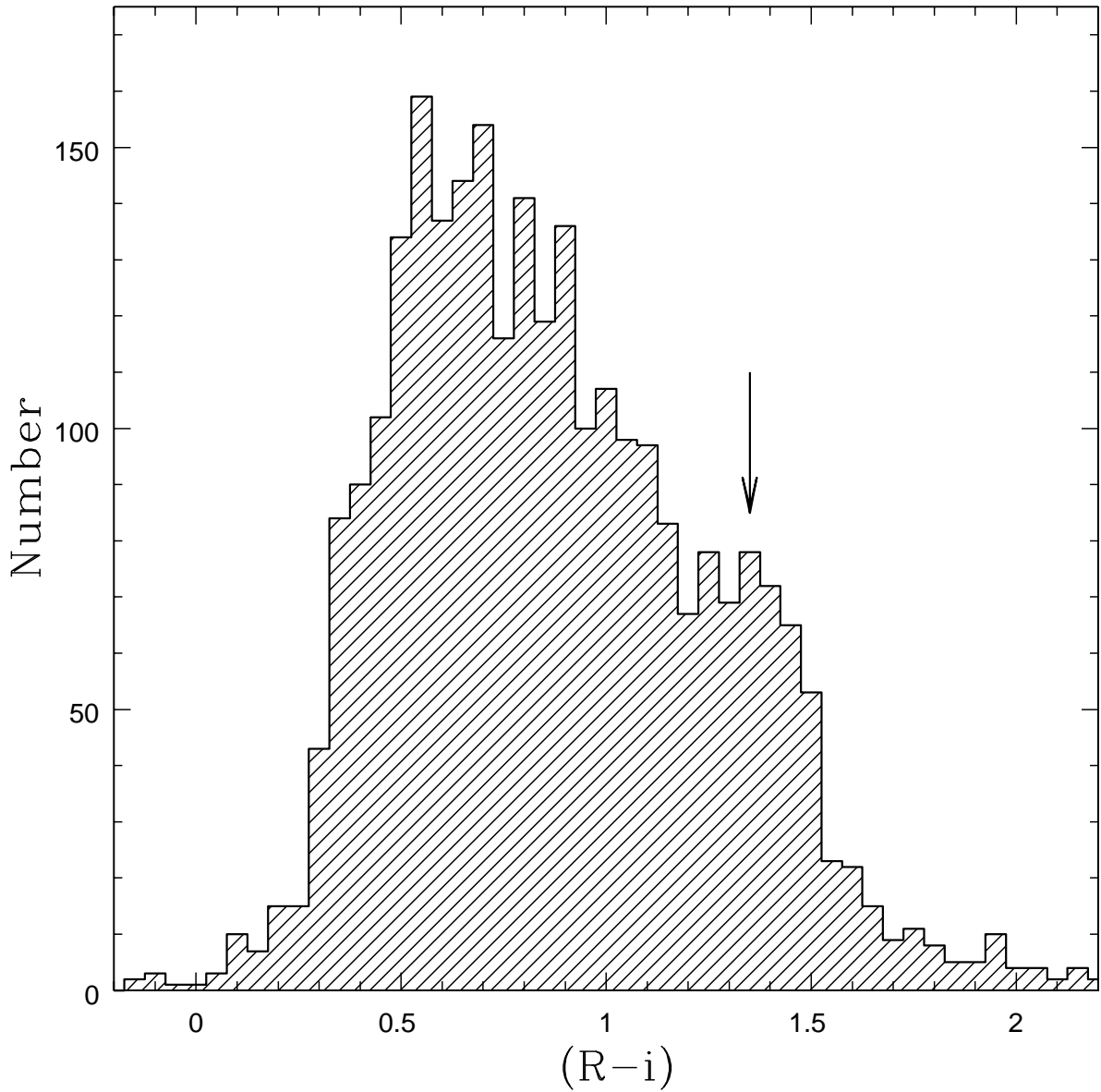}
\caption{Histogram of the $(R-i)$ colors of all galaxies with $20 \le
i \le 23.5$. The vast majority of galaxies belong to the field
population which has an average color of $(R-i) \approx 0.7$.  The red
sequence of supercluster members is defined by the narrower peak at
$(R-I) \approx 1.4$ (indicated by the arrow) and is easily
distinguished from the distribution of field galaxies. Fitting two
Gaussians functions to this distribution, we find that $\sigma_1 =
0.30$ mag for the field population and $\sigma_2 = 0.15$ mag for the
red sequence.}
\label{hist}
\end{figure}

With the multi-band imaging, we have generated photometry on a
complete sample of galaxies in a contiguous area of $10.3^{'} \times
18.3^{'}$ or $3.1~h^{-1}~{\rm Mpc} \times 5.5~h^{-1}~{\rm Mpc}$ at the
supercluster redshift of $z \approx 0.91$. We show the resulting
$(B-R)$ and $(R-i)$ color-magnitude (CM) diagrams in Figure~\ref{cm}.
In both diagrams, we see a well-defined color-magnitude sequence which
is redder than the vast majority of galaxies which comprise the field
population.  This red sequence of galaxies is considerably tighter in
the $(R-i$) CM diagram where it is observed at $(R-i) \approx 1.4$. In
the $(B-R)$ CM diagram, the larger color scatter in this red sequence
is a result of the fact that many of these galaxies lie at or beyond
the completeness limits of this survey.  Figure~\ref{hist} shows a
histogram of $(R-i)$ colors.  In this figure, the red sequence of
galaxies can be clearly distinguished from the field population where
it is a red peak superimposed on the large distribution of bluer field
galaxies.  Fitting two Gaussian functions to this distribution, we
find that the standard deviation of the red peak is 0.15 mag.

A tight, red color-magnitude relation is typical of the central regions of
massive clusters both in the local universe and at intermediate and high
redshift (e.g., Dressler 1980; Butcher \& Oemler 1984; Stanford, Dickinson \&
Eisenhardt 1995, 1997).  The galaxies contained in this ``red locus'' are the
elliptical and S0 galaxies which comprise the majority of the cluster
population.  The early-type galaxies are characterized by their red color and
their small color scatter, typically less than 0.2 mag.  Studies of clusters
from $z \sim 1$ to the present epoch imply that the observed color trend in
this red envelope of galaxies is consistent with passive evolution of an old
stellar population formed in a relatively synchronized burst of star formation
at $z \ga 2$ (e.g., Ellis et al.\ 1997; Stanford et al.\ 1995, 1997; Bower,
Kodama \& Terlevich 1998).

\begin{figure}
\plotone{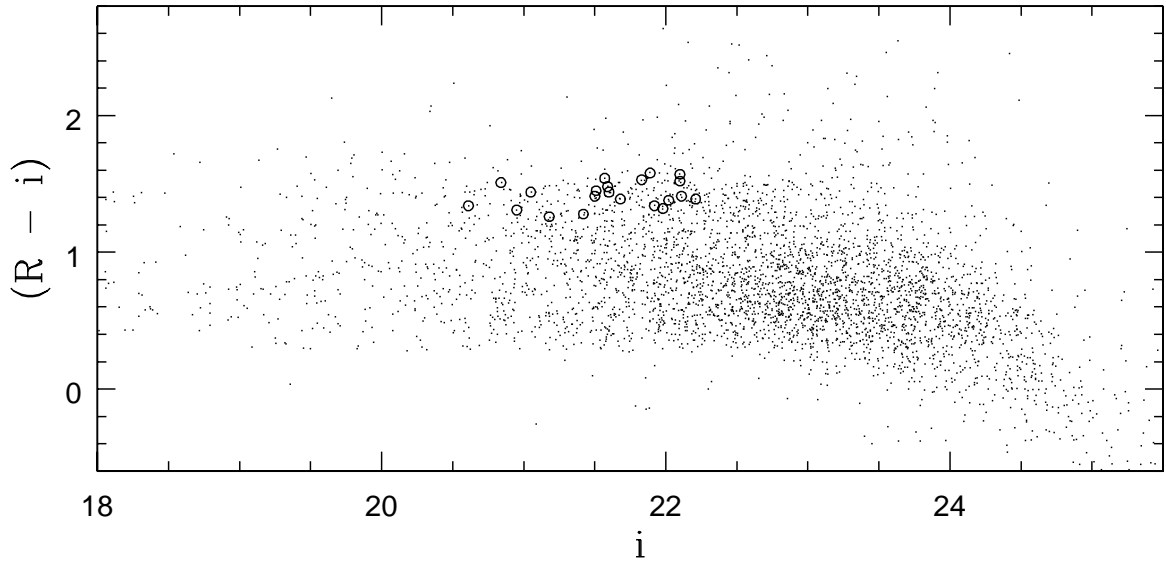}
\caption{The same $(R-i)$ color--magnitude diagram as shown in
Figure~\ref{cm}. The circles indicate all of the
spectroscopically-confirmed, early-type galaxies in the two clusters
which lie within the field-of-view of our supercluster observations.
The color of these galaxies are consistent with the color of the red
locus, indicating that there exists a large population of red,
early-type galaxies at the supercluster redshift.}
\label{ri}
\end{figure}

We have confirmed that the red sequence observed in our supercluster
field corresponds to a population of old, early-type galaxies. Within
this field, there are 21 galaxies which have an absorption spectrum
which is typical of an early-type galaxy and are
spectroscopically-confirmed members of either CL1604+4304 at $z =
0.896$ or CL1604+4321 at $z = 0.924$ (Postman et al.\ 1998, 2000). In
Figure~\ref{ri}, we indicate those galaxies on the $(R-i)$ CM
diagram. All of these galaxies fall directly on the observed red
locus, confirming that it is comprised mainly of early-type galaxies
at the supercluster redshift of $z \approx 0.91$.  As discussed in
Postman et al.\ (1998, 2000), the color of these galaxies are
consistent with the passive evolution of a stellar population which
formed at redshifts of $z \ga 2-5$.  While the presence of a red
sequence of early-type galaxies is typical of the central
$0.5~h^{-1}~{\rm Mpc}$ in massive clusters, the fact that we observe
such a well-defined red sequence over the entire $5.5~h^{-1}~{\rm
Mpc}$ scale of the supercluster strongly supports the existence of a
large scale structure at $z \approx 0.91$.

\subsection{The Spatial Distribution}

\begin{figure}
\epsscale{0.65}
\plotone{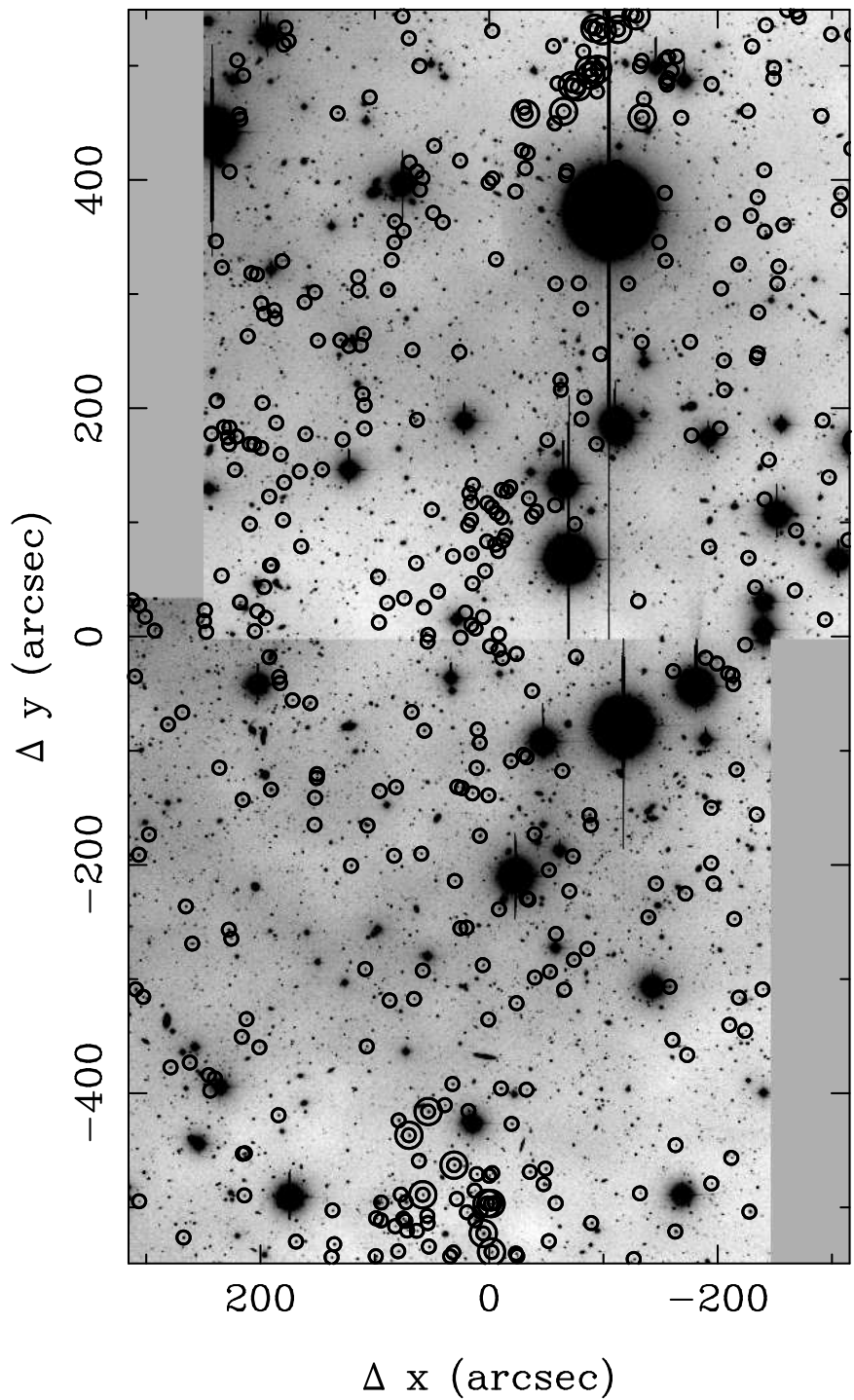}
\caption{Composite $i$ band image of the supercluster field (North is
$\uparrow$ and East is $\rightarrow$). The cluster centers of
CL1604+4304 and CL1604+4321 are at the bottom and top, respectively,
of this image.  Large circles indicate those confirmed cluster members
with spectra that are consistent with early-type galaxies.  The likely
supercluster members based on our color selection (see \S 3.2) are
indicated by small circles. Their spatial distribution clearly shows
the presence of a large scale system of galaxies encompassing the two
rich clusters, as well as a new cluster concentration in the center of
the field at the coordinates $(0,+100)$.}
\label{pal}
\end{figure}

Based on the $(R-i$) color-magnitude sequence, we can select galaxies
which are likely supercluster members. We have chosen all galaxies
which form the red locus ($1.2 \le R-i \le 1.7$) and can be
spectroscopically observed in a reasonable time on a 10-m class
telescope ($20 \le i \le 23$).  The resulting sample contains 418
galaxies which are shown on the composite $i$ band image in
Figure~\ref{pal}.  The clusters CL1604+4304 and CL1604+4321 are at the
bottom and top, respectively, of this image.  The spatial distribution
of red galaxies clearly delineates the large scale system of galaxies
which encompasses the two rich clusters. These galaxies are spread
throughout the full field, being noticeably more concentrated near the
cluster centers. In addition, we observe a strong concentration of red
galaxies directly between CL1604+4304 and CL1604+4321 at
$16~04~25.7~+43~14~44.7~{\rm (J2000)}$.  Based on a control field, we
find the number density of red field galaxies, as defined by our color
selection, is $0.7 \pm 0.1$ galaxies per arcmin$^{2}$.  Within a
radius of $0.2~h^{-1}~{\rm Mpc}$, the new concentration is overdense
in red galaxies by a factor of $\sim 20$ compared to this control
field. This overdensity is equivalent to that observed in the two
original clusters; CL1604+4304 and CL1604+4321 are overdense by a
factor of 20 and 13, respectively.  At half an Abell radius, the
overdensity of red galaxies in the three clusters is a factor of $\sim
5 - 6$. These data clearly indicate that we have detected a third,
massive cluster associated with this large scale structure. We also
observe at least one other, although more marginal, overdensity in
this field, suggesting that this supercluster may contain additional
clusters beyond the three discussed here. The spatial distribution of
red galaxies provides further confirmation of a large scale structure
spanning at least $5.5~h^{-1}$ Mpc.

\section{Conclusions}

We have confirmed the existence of a supercluster at $z = 0.91$ with
deep, multiband imaging taken at the Palomar 200-in telescope. In the
resulting color--magnitude diagrams, we find a relatively tight,
color--magnitude sequence of red galaxies.  The characteristic color
of this sequence corresponds directly to the colors of the confirmed
early-type galaxies in the two member clusters, CL1604+4304 and
CL1604+4321.  Therefore, we have identified a large population of
early-type galaxies within the redshift range of $z \sim 0.89 - 0.93$.
These red galaxies cover the entire $5.5~h^{-1}~{\rm Mpc}$ region
which separates the two member clusters. They delineate the full
extent of the large scale structure, while clearly encompassing the
two clusters.  Based on the distribution of red galaxies over this
field, we have identified another rich cluster associated with this
system. It lies directly between CL1604+4304 and CL1604+4321, and its
overdensity of red galaxies is approximately equal to that of the two
originally identified clusters.  The strong red sequence of galaxies
and their distribution on the sky leave no doubt that there exists a
large scale structure at $z \approx 0.91$. This supercluster is the
first massive structure confirmed at such a high redshift. At higher
redshift, there is a two cluster system at $z = 1.27$ (Rosati et al.\
1999). The nature of large scale structure surrounding this pair has
not yet been explored.

In order to study this structure even further, we are planning to
perform two additional observational studies.  Firstly, we will
complete an extensive spectroscopic survey using the multislit
capability of the Low Resolution Imaging Spectrograph (LRIS; Oke et
al.\ 1995) on the Keck 10-m telescope. Based on our accurate color
selection, we expect to measure redshifts for over 400 supercluster
members.  This sample is comparable in size to the largest
spectroscopic studies of the Shapley and Corona Borealis superclusters
(Quintana et al.\ 1995; Small et al.\ 1998).  Secondly, we will
perform multi-band imaging of the entire supercluster region,
including two flanking fields, using the Large Field Camera (LFC;
Metzger et al.\ 2000) on the Palomar 200-in telescope.  This survey
will provide $u'g'r'i'z'$ photometry for over 10,000 galaxies which
will be used to calculate photometric redshifts accurate to $\Delta z
= 0.06$ (Brunner, Connolly \& Szalay 1999).  These data will allow us
to study galaxy properties as a function of position and local density
within the supercluster; to measure the mass distribution on
intermediate scales and estimate $\Omega_{o}$; and to examine the
early stages of cluster formation through the accretion of matter in
the cluster infall regions. As a result, this high-redshift structure
will be one of the most well-studied superclusters.

\vskip 0.3cm 

We thank the anonymous referee for useful comments. LML is supported
by NASA through Hubble Fellowship grant HF-01095.01-97A from the Space
Telescope Science Institute, which is operated by the Association of
Universities for Research in Astronomy, Inc., under NASA contract NAS
5-26555.

\newpage 

\begin{thebibliography}{}

\bibitem[]{} Bahcall. N.A. \& Soneira, R.M. 1984, \apj, 277, 27
\bibitem[]{} Bertin, E. \& Arnouts, S. 1996, \aap, 117, 393
\bibitem[]{} Bower, R.G., Kodama, T., \& Terlevich, A. 1998,
\mnras, 299, 1193
\bibitem[]{} Butcher, H. \& Oemler, A. 1978, \apj, 226, 559
\bibitem[]{} Brunner, R.J, Connolly, A.J., \& Szalay, A.S. 1999,
\apj, 516, 563
\bibitem[]{} Cen, R. 1994, \apj, 424, 22 
\bibitem[]{} Davis, M., Tonry, J., Huchra, J., \& Latham, D.W. 1980,
\apj, 238, 11
\bibitem[]{} Dressler, A. 1980, \apjs, 42, 565
\bibitem[]{} Ellis, R.S., Smail, I., Dressler, A., Couch, W.J.,
  Omeler, A., Butcher, H., \& Sharples, R. 1997, \apj, 483, 582
\bibitem[]{} Hoffman, Y., Shaham, J., \& Shaviv, G.\ 1982,
\apj, 262, 413
\bibitem[]{} Kaiser, N., Wilson, G., Luppino, G., Kofman, L., Gioia,
I., Metzger, M., \& Dahle, H. 2000, \apj, in press
\bibitem[]{} Kells, W., Dressler, A., Sivaramakrishnan, A., Carr, D.,
Koch, E., Epps, H., Hilyard, D., \& Pardeilhan, G. 1998, \pasp, 110,
1487
\bibitem[]{} Landolt, A. U. 1992, \aj, 104, 340
\bibitem[]{} Lubin, L.M., Postman, M., Oke, J.B., Ratnatunga, K.U.,
  Gunn, J.E., Hoessel, J.G., \& Schneider, D.P.  1998, \aj, 116, 584
\bibitem[]{} Lubin, L.M., Postman, M., Oke, J.B., Brunner, R., Gunn,
J.E., \& Schneider, D.P. 2000, \aj, in preparation
\bibitem[]{} Metzger, M.R. et al.\ 2000, \pasp, in preparation
\bibitem[]{} Oke, J.B. et al. 1995, \pasp, 107, 375
\bibitem[]{} Oke, J.B., Postman, M., \& Lubin, L.M. 1998, \aj, 116,
549
\bibitem[]{} Peebles, P.J.E. 1986, \nat, 321, 27
\bibitem[]{} Postman, M., Lubin, L.M., \& Oke, J.B. 1998, \aj, 116, 560
\bibitem[]{} Postman, M., Lubin, L.M., \& Oke, J.B. 2000, \aj, in preparation
\bibitem[]{} Postman, M., Geller, M.J., \& Huchra, J.P. 1988,
\apj, 95, 267
\bibitem[]{} Quintana, H., Ramirez, A., Melnick, J., Raychaudhury, S.,
\& Slezak, E. 1995, \aj, 110, 463
\bibitem[]{} Rosati, P., Stanford, S.A, Eisenhardt, P.R., Elston, R.,
Spinrad, H., Stern, D., \& Dey, A. 1999, \aj, 118, 76
\bibitem[]{} Shaya, E.J. 1984, \apj, 280, 470
\bibitem[]{} Small, T.A., Chung-Pei, M., Sargent, W.L.W., \& Hamilton,
D. 1998, \apj, 492, 45
\bibitem[]{} Stanford, S.A., Eisenhardt, P.R.M., \& Dickinson, M. 
  1995, \apj, 450, 512
\bibitem[]{} Stanford, S.A., Eisenhardt, P.R.M., \& Dickinson, M. 
  1997, \apj, 492, 461
\bibitem[]{} Szalay, A.S., Connolly, A.J., \& Szokoly, G.P. 1998,
\aj, 117, 68

\end{thebibliography}
\end{document}